# A 49.8mm$^2$ Fully Integrated, 1.5m Transmission-Range, High-Data-Rate IR-UWB Transmitter for Brain Implants

Cong Ding, Mingxiang Gao, Anja K. Skrivervik, Mahsa Shoaran

EPFL, Lausanne, Switzerland

Implantable neural interfaces hold great potential to restore function in individuals impaired by paralysis, limb amputations, or various neurological conditions. The need for precise mapping of neuronal activity across various brain regions and enhancing the information transfer rate has driven a significant increase in the number of recording channels, with recent systems incorporating thousands or more [1-2]. This necessitates wireless links capable of handling throughputs in the range of hundreds of Mb/s, posing a significant challenge in terms of power consumption, size, and transmission range for wireless implants. Body channel communication (BCC) has seen a rise in adoption for brain implants owing to its ability to achieve millimeter-scale form factors [3-4]. Yet, it faces limitations in both data rate and transmission distance. Alternately, Impulse-Radio Ultra-Wideband (IR-UWB) communication offers a promising solution thanks to its high data rate and low power consumption [5-6]. However, existing IR-UWB transmitters (TXs) are hindered by their cm-level transmission range and large dimensions, making them suboptimal for chronic implantation. Far-field RF radiation achieving m-level transmission distances offers substantial freedom of movement for patients. However, it demands an efficient wireless link that conforms to strict power consumption requirements of tens of mW/cm$^2$ for the brain. To address the challenge of extending the transmission range of implantable TXs while also minimizing their size and power consumption, this paper introduces a transcutaneous, high data-rate, fully integrated IR-UWB transmitter that employs a novel co-designed power amplifier (PA) and antenna interface for enhanced performance. With the co-designed interface, we achieved the smallest footprint of 49.8mm$^2$ (8.3mm×6mm) and the longest transmission range of 1.5m compared to the state-of-the-art IR-UWB TXs [5-6].

Figure 1 presents the architecture of the proposed TX which incorporates an on-off-keying (OOK) modulation scheme combined with phase shift keying (PSK)-based scrambling. The utilization of PSK scrambling offers enhanced control over polarity, thus effectively removing discrete spectral tones in the OOK output spectrum to comply with FCC regulatory requirements. The quadrature local oscillator (LO) signals are generated by a 2-stage ring oscillator (RO)-based integer-N wide-band phase-locked loop (PLL), providing an LC-VCO-like jitter performance. The pulse generator outputs a 2ns-pulse-width OOK data, which is fed into a pulse shaper (PS) with programmable delay line (DL). The PS, in conjunction with the switched-capacitor PA (SCPA), conducts FIR filtering in the RF domain, thereby enhancing spectral efficiency. The wireless link is established by an off-chip dipole antenna, selected for its compatibility with miniaturized implants as it does not require a large ground plane in contrast to monopole antennas.

Figure 2 presents the block diagrams of the inverter-based phase multiplexer (PHMUX), PS, and the SCPA. A fully differential architecture is utilized for both PHMUX and SCPA, eliminating the need for an off-chip balun. To enhance the power and area efficiency while ensuring effective sidelobe suppression, a 4-bit triangular template is employed. This template can be configured as either symmetrical or asymmetrical, thereby improving inter-symbol interference (ISI) performance. Figure 2 (top-right) compares the simulated output spectrum for the proposed modulation scheme against the ideal triangular envelope, demonstrating a comparable performance in sidelobe suppression and bandwidth of the main lobe.

Figure 3 illustrates the circuit implementation of the digital/voltage-controlled RO, featuring a pair of delay elements and hybrid-controlled resistors. The oscillation frequency is controlled by the 4-bit digital control word (FC) to overcome PVT variations, along with two analog signals generated by the differential loop filter (i.e., VCP and VCN). To minimize substrate noise coupling, we employ a differential charge pump (CP) and loop-pass filter (LPF), resulting in an almost twofold increase in tuning range compared to the single-ended configuration. The measured PLL locking frequency ranges from 4GHz to 6GHz under a 1.2V supply. Using a 350MHz reference clock, the measured phase noise performance exhibits in-band phase noise of -80dBc/Hz at 1MHz and out-of-band phase noise of -101dBc/Hz at 100MHz offset. The double sideband (DSB) rms jitter of the PLL within the 10kHz to 100MHz range was measured to be 18.9ps at 5.6GHz.

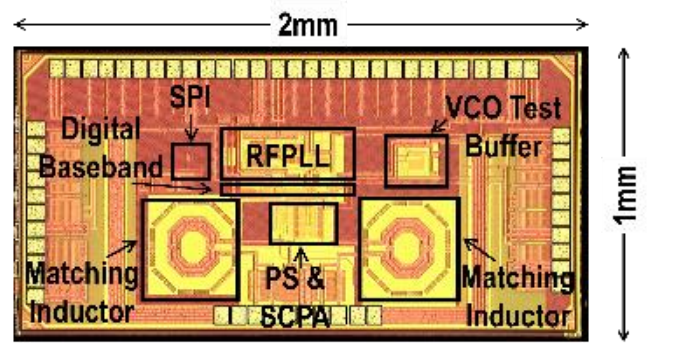

Die micrograph.

The SCPA employs a capacitive combination of two digitally generated impulses with out-of-phase RF and in-phase baseband components (Fig. 2). Thus, the latter is effectively canceled, minimizing common-mode voltage fluctuations that could otherwise induce spectrum leakage at low frequencies. A 14mm$^2$ meander dipole antenna was characterized in-vitro through subcutaneous implantation into a chicken phantom, as illustrated in Fig. 4. The measured far-field radiation pattern exhibits a maximum gain of -14.6dBi in the main lobe at 4.6GHz. We measured the radiation pattern to determine the optimal placement of the external RX, aiming to maximize the link efficiency in the direction of the main lobe. The 3-dB bandwidth of the wireless link reaches 800MHz between the implantable antenna and an external UWB antenna. The input impedance of the dipole antenna is designed to match the optimum load impedance of the proposed SCPA around 4.5GHz, as depicted in Fig. 4. This approach eliminates the need for an off-chip matching network, further simplifying implementation and enhancing the energy efficiency of the implantable TX.

A prototype IR-UWB transmitter was implemented in 65nm CMOS. The total chip area including the pads is 2mm×1mm. Figure 5 shows the wireless measurement setup. The PA's output is measured through connection to an oscilloscope via a twisted pair cable and a chip balun. Excluding the 1.5-dB power loss caused by this setup, the PA achieves a maximum output power of around 1.4dBm. We measured a maximum OOK modulation rate of 800Mb/s. The transient output waveform of the PA, featuring the proposed asymmetrical envelope and PSK scrambling, is depicted in Fig. 5. We conducted in-vitro measurements of the wireless link by placing the wireless TX module at varying distances from the RX. The RX includes an 8.5dBi UWB antenna, an RF filter, a 1.2dB-NF LNA, and a high-speed oscilloscope serving as an ADC. Given the 800MHz bandwidth limitation of the wireless link, we achieved a 500Mb/s data rate. Fig. 5 displays RX received signals at various TX-RX distances and data rates. Demodulation is performed using MATLAB, achieving a BER below $10^{-4}$. To the best of our knowledge, this design achieves the smallest footprint while enabling a 100-cm and 150-cm transmission range at 500Mb/s and 375Mb/s for brain implants, respectively.

Figure 6 presents a comparison table with the state-of-the-art TX designs for brain implants. Since OOK modulation does not require a precise carrier, our TX's power consumption can be further reduced by utilizing an open-loop RO instead of relying on the PLL as the LO. In comparison to the state-of-the-art, our TX features the smallest footprint, a meter-level transmission range, and one of the lowest normalized energy efficiencies.

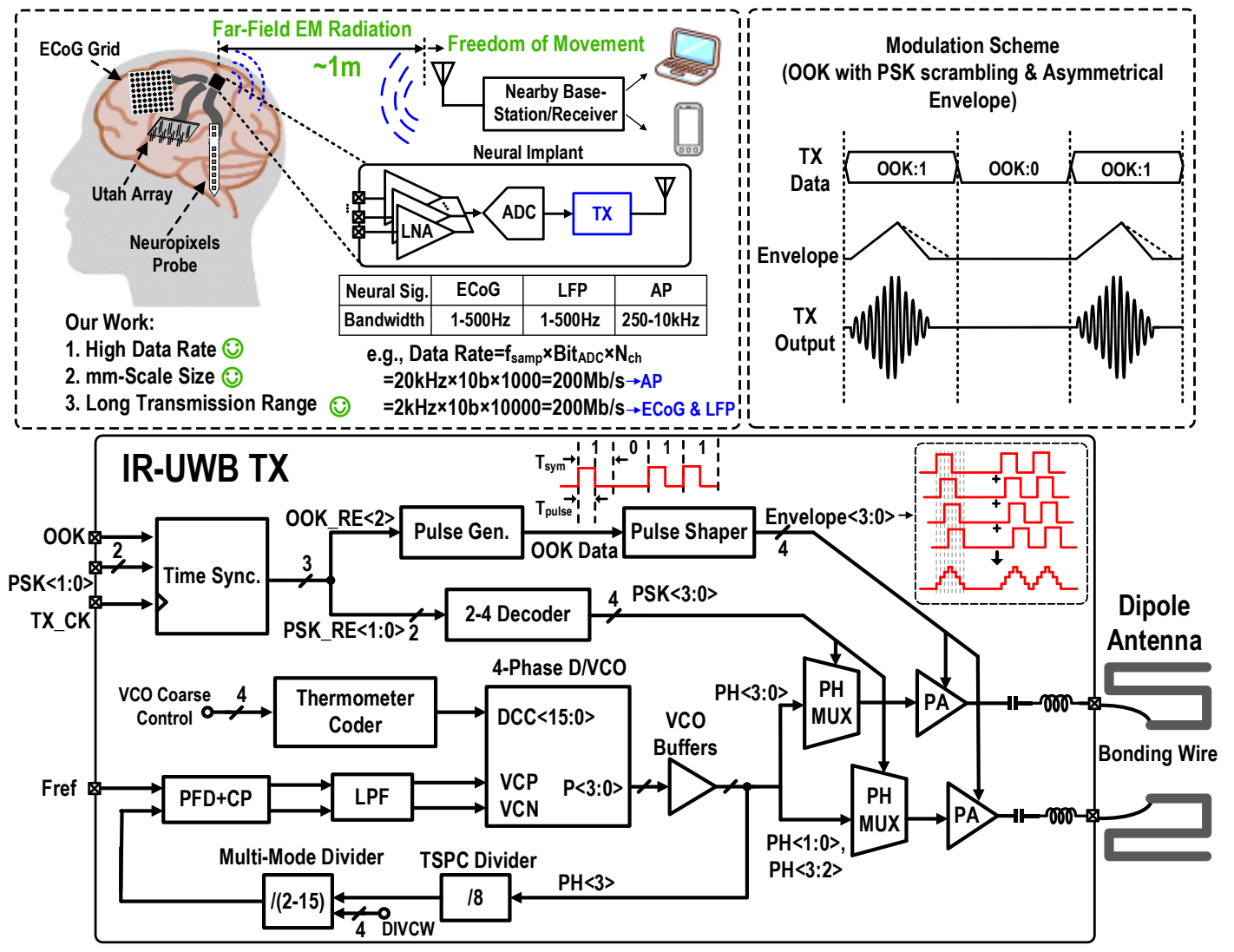

Fig. 1. The motivation, modulation scheme, and block diagram of the proposed IR-UWB transmitter.

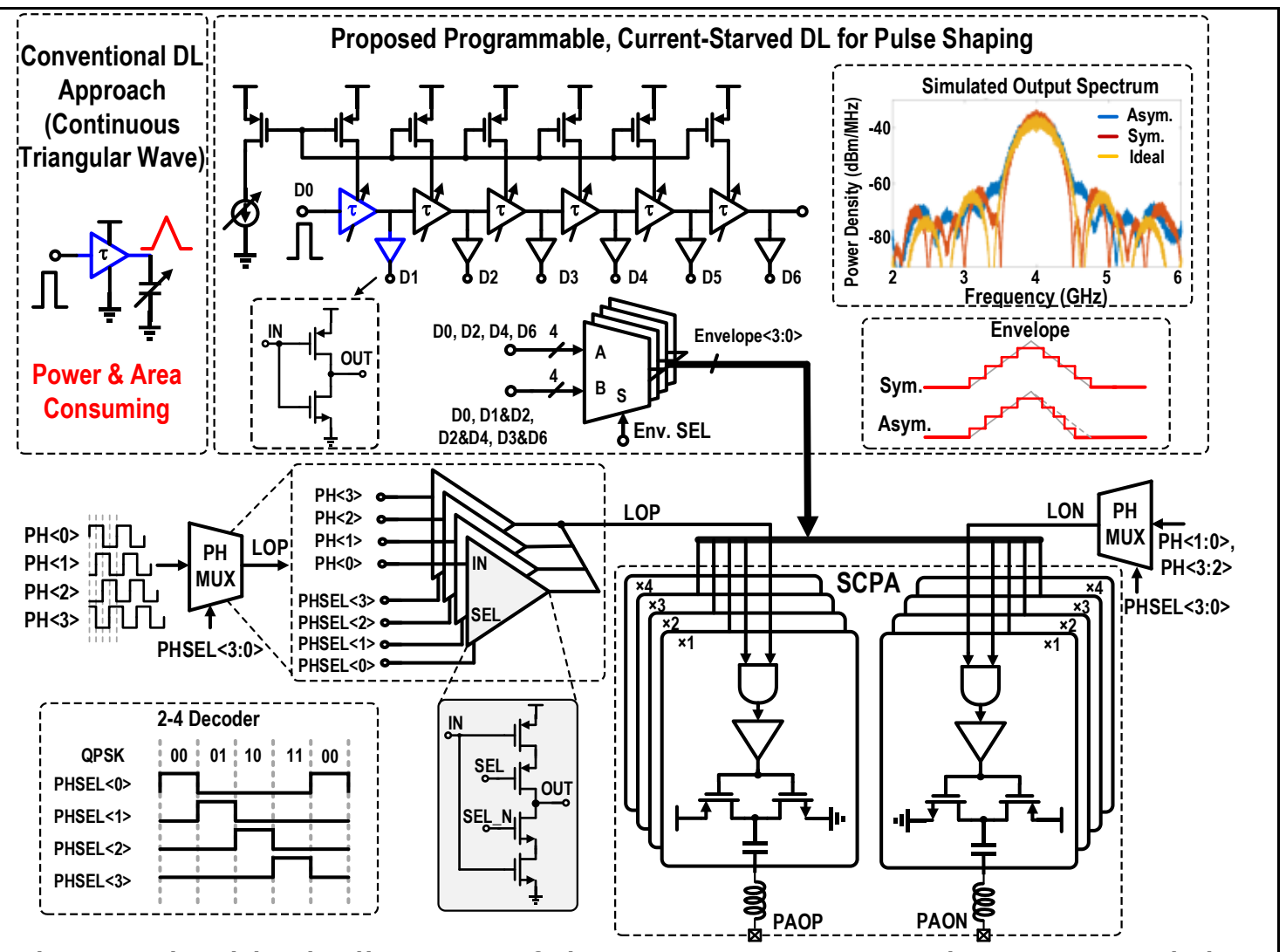

Fig. 2. The block diagram of the PS, PHMUX, and SCPA, and the simulated output spectrum of the TX with the proposed modulation scheme against triangular envelope in MATLAB.

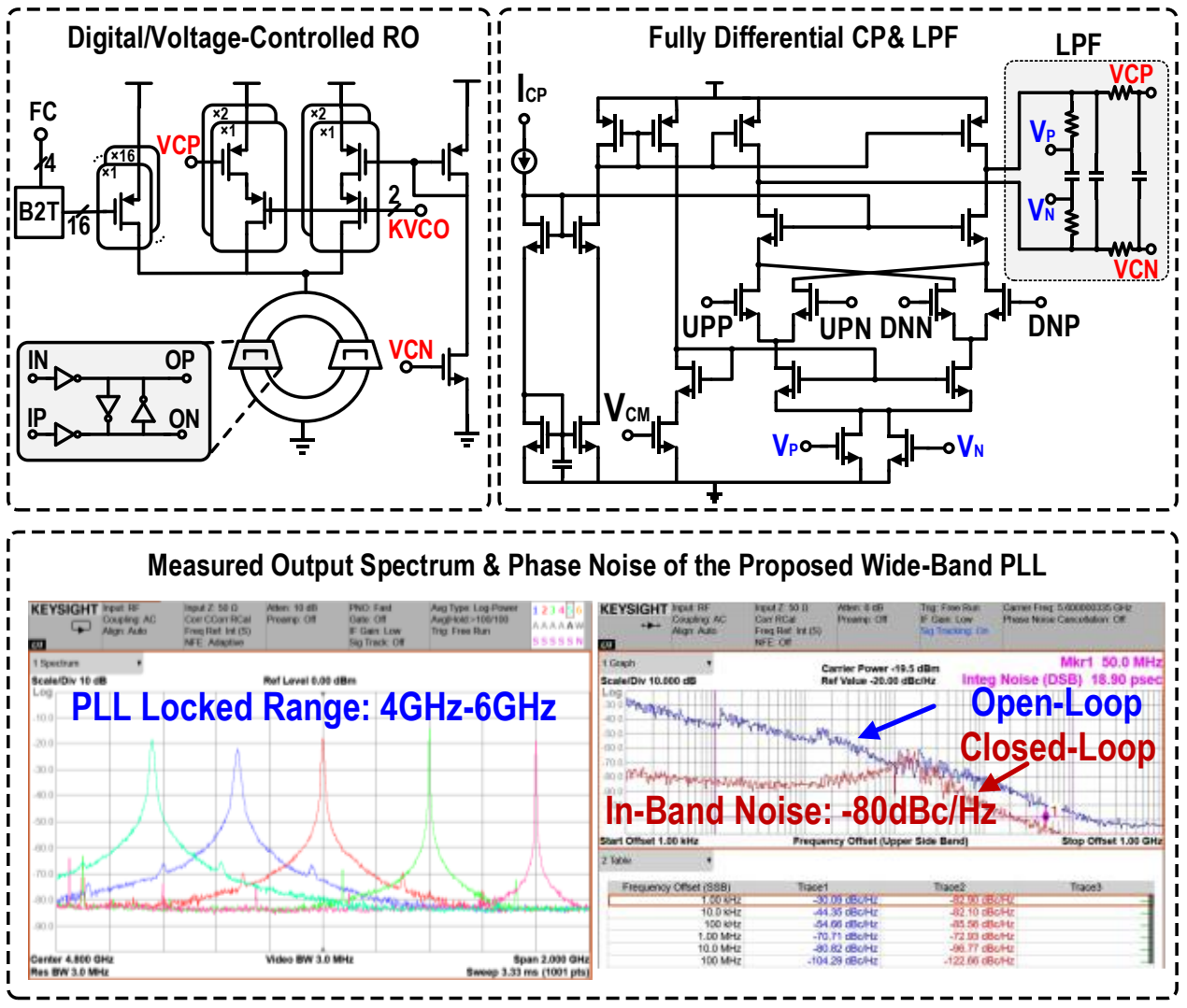

Fig. 3. Circuit implementations of CP, LPF, and RO, and the measured output spectrum and phase noise of the PLL.

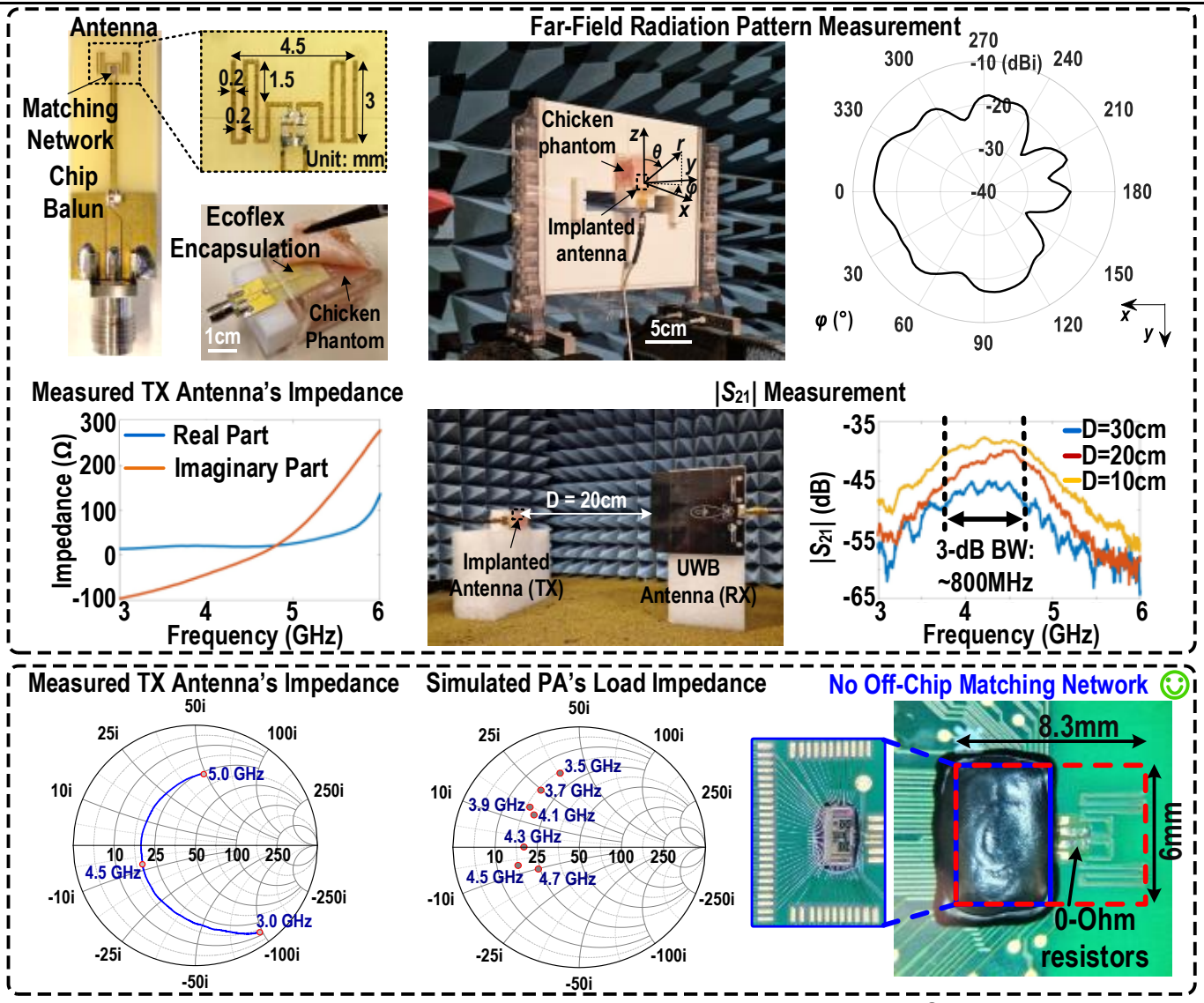

Fig. 4. The characterization of the fabricated 14mm$^2$ dipole antenna, and details of the antenna/PA co-design.

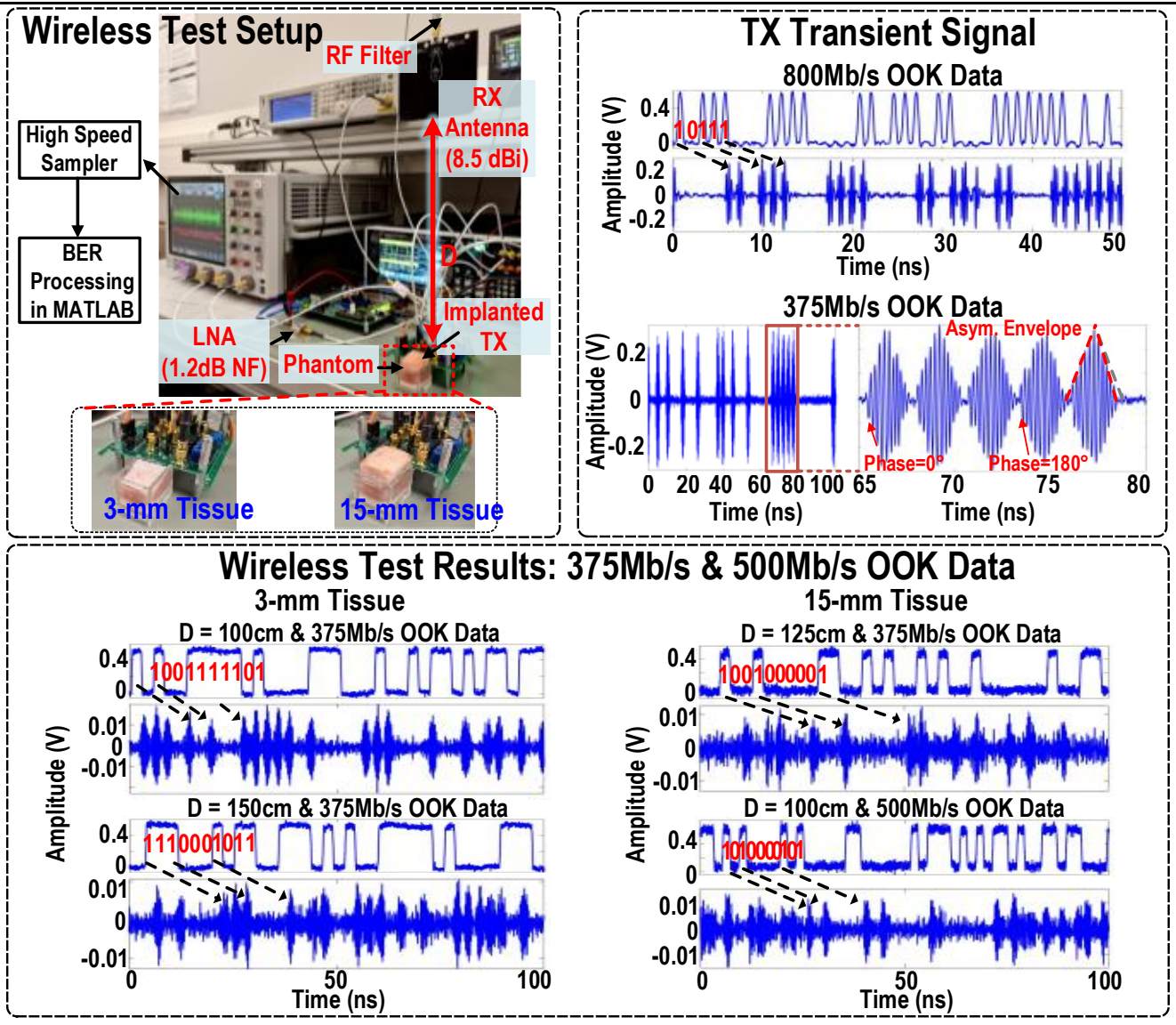

Fig. 5. The wireless test setup and measured transient output waveforms.

| Parameter | VLSI'21 [3] | TMTT'22 [4] | ISSCC'22 [5] | ISSCC'23 [6] | This Work |
|---|---|---|---|---|---|
| Process (nm) | 65 | 40 | 28 | 40 | 65 |
| Link Method | BCC | BCC | RF | RF | RF |
| Frequency Band (GHz) | - | - | 6-9 | 3.1-5 | 3.1-6 |
| Architecture | Galvanic Coup. | Galvanic Coup. | Up-Conversion | Edge Combine | Up-Conversion |
| Modulation Scheme | OOK | OOK | 4PPM+8PSK+4 PAM | D16PPM+PWK+ DBPSK | OOK w Asym. Envelope |
| Data Rate (Mb/s) | 0.006-10 | >250 | 1660 | 1800 | 100-800 |
| PA Max. Output Power (dBm) | - | - | −8 | −7.1 | 1.4 |
| TX Power (mW) | 0.52 | 0.5 | 9.69 | 4.09 | 13.2 |
| Energy Eff. (pJ/b) | 52@10Mb/s | 2 | 5.8 | 2.3 | 16.5 |
| TX Antenna Size (mm$^2$) | - | - | 65.8 | 100* | 14 (Co-Designed) |
| TX Antenna Input Impedance (Ω) | - | - | 50 | 50 | 18.9−i8.4 (PA Load Line Impedance) |
| Implant Size (mm$^2$) | 11.25 | 26 | 105 | - | 49.8 |
| Tissue Thickness (mm) (In-Vitro) | Diam. 11-cm PBS | 10 (Muscle) | 15 (Skin, Fat) | 18 (Skin, Fat) | 15 (Skin, Fat) |
| TX Antenna Gain (dBi) | - | - | −8.5 | 2.6 | -19.2 (Incl. Tissue Loss) |
| RX Antenna Gain (dBi) | - | - | 5 | 3.5 | 8.5 |
| Max. Distance (cm) | 5.5 | 1 | 2 | 15 | 100@500Mb/s‡ |
| Normalized Energy Eff. (pJ/b/m) [5] | 945 | 200 | 45@1.43Gb/s 290@1.66Gb/s | 15.3* | 26.4@500Mb/s‡ |

* Using a high-gain commercial antenna

‡ Limited by the Antenna bandwidth

Fig. 6. Comparison to the state-of-the-art TX designs.